\documentclass[aps,prb,twocolumn,showpacs,floatfix,groupedaddress
] {revtex4-1}
\usepackage[T1]{fontenc}
\usepackage[latin9]{inputenc}

\usepackage{graphicx}
\usepackage{amssymb}
\usepackage{amsmath}
\usepackage{color}



\begin{document}

\newcommand{\be}{\begin{equation}}
\newcommand{\ee}{  \end{equation}}
\newcommand{\ba}{\begin{eqnarray}}
\newcommand{\ea}{  \end{eqnarray}}
\newcommand{\ve}{\varepsilon}

\newcommand{\WB}[1]{\textbf{WB: #1}}

\title{Shot noise of charge and spin transport in a junction with a precessing molecular spin}

\author{Milena Filipovi\'{c}}
\author{Wolfgang Belzig}
\affiliation{Fachbereich Physik, Universit\"at Konstanz, D-78457 Konstanz, Germany}

\date{\today}
\begin{abstract}
Magnetic molecules and nanomagnets can be used to influence the electronic transport in mesoscopic junction. In a magnetic field the precessional motion leads to resonances in the dc- and ac-transport properties of a nanocontact, in which the electrons are coupled to the precession. Quantities such as the dc conductance or the ac response provide valuable information, such as the level structure and the coupling parameters. Here, we address the current-noise properties of such contacts. This encompasses the charge current and spin-torque shot noise, which both show a steplike behavior as functions of bias voltage and magnetic field. The charge-current noise shows pronounced dips around the steps, which we trace back to interference effects of electrons in quasienergy levels coupled by the molecular spin precession. We show that some components of the noise of the spin-torque currents are directly related to the Gilbert damping and hence are experimentally accessible. Our results show that the noise characteristics allow us to investigate in more detail the coherence of spin transport in contacts containing magnetic molecules.
\end{abstract}
%\pacs{73.23.-b, 75.76.+j, 85.65.+h, 85.75.-d}
\maketitle

\section{Introduction}

Shot noise of charge current has become an active research topic in recent decades, since it enables the investigation of microscopic transport properties, which cannot be obtained from the charge current or conductance.\cite{ButtikerBlanter2000}
It has been demonstrated that spin-flip induced fluctuations in diffusive conductors connected to ferromagnetic leads enhance the noise power, approaching the Poissonian value. \cite{Mishchenko2003,Belzig2004} 
Accordingly, the Fano factor defined as $F=S(0)/e\lvert I \rvert$, which describes the deviation of the shot noise from the average charge current, equals 1 in this case.
On the other hand, it has been shown that shot noise in a ferromagnet-quantum-dot-ferromagnet system with antiparallel magnetization alignments can be suppressed due to spin flip, with $F<1/2$. \cite{SouzaJauho2008}

The quantum-interference phenomenon, which is a manifestation of the wave nature of electrons, has attracted a lot of attention. The quantum-interference effects occur between coherent electron waves in nanoscale junctions. \cite{MiroshnichenkoFlach2010} Quantum interference in molecular junctions influences their electronic properties. \cite{GuedonValkenier2012,VazquezSkouta2012,Stadler2009,Ratner19904877,HansenSolomon2009} The Fano effect\cite{Fano1961} due to the interference between a discrete state and the continuum has an important role in investigation of the interference effects in nanojunctions, which behave in an analogous way, and are manifested in the conductance or noise spectra. \cite{MiroshnichenkoFlach2010,FaistCapasso1997,Entin-Wohlman2007} Particularly interesting examples involve spin-flip processes, such as in the presence of Rashba spin-orbit interaction,\cite{SanchezSierra2006,PiotrStefanski2010} a rotating magnetic field,\cite{Floquet4} or in the case of the magnetotransport.\cite{GurvitzMozyrsky2005,GurvitzIEEE,MozyrskyFedichkin2002}

In the domain of spin transport it is interesting to investigate the noise properties, as the discrete nature of electron spin leads to the correlations between spin-carrying particles. The spin current is usually a nonconserved quantity that is difficult to measure, and its shot noise depends on spin-flip processes leading to spin-current correlations with opposite spins. \cite{Lamacraft2004,ZareyanBelzig2005,WangWang2004} 
The investigation of the spin-dependent scattering, spin accumulation,\cite{Ouyang2008} and attractive or repulsive interactions in mesoscopic systems can be obtained using the shot noise of spin current,\cite{Sauret2004} as well as measuring the spin relaxation time. \cite{Lamacraft2004,Sauret2004} 
%One should mention that 
Even in the absence of charge current, a nonzero spin current and its noise can still emerge. \cite{StevensSmirl2003,BrataasTserkovnyak2002,WangWang2004} Several works have studied the shot noise of a spin current using, e.g., the nonequilibrium Green's functions method and scattering matrix theory.\cite{Dragomirova2007,WangWang2004,Yhedhourhan2007,YuZhan2013}

It was demonstrated that the magnetization noise originates from
transferred spin current noise via a fluctuating spin-transfer torque
in ferromagnetic-normal-ferromagnetic systems,\cite{ForosBrataas2005}
and magnetic tunnel junctions.\cite{Chudnovsky2008} Experimentally,
Spin Hall noise measurments have been demonstrated,\cite{Kamra2014}
and in a similar fashion the spin-current shot noise due to magnon
currents can be related to the nonquantized spin of interacting
magnons in ferri-, ferro-, and antiferromagnets.\cite{Kamra2016,Kamra2017}
Quantum noise generated from the scatterings between the magnetization
of a nanomagnet and spin-polarized electrons has been studied theoretically as
well.\cite{WangSham2012,WangSham2013} 
The shot noise of spin-transfer torque was studied recently using a magnetic quantum dot connected to two 
noncollinear magnetic contacts.\cite{YuZhan2013} According to the
definition of the spin-transfer torque,\cite{slonczewski,berger} both
autocorrelations and cross-correlations of the spin-current
components contribute to the spin-torque noise.  

In this article, we study theoretically the noise of charge and spin
currents and spin-transfer torque in a junction connected to two
normal metallic leads. The
transport occurs via a single electron energy level interacting with
a molecular magnet in a constant magnetic field. The spin of the
molecular magnet precesses 
around the magnetic field with the Larmor frequency, which is 
kept undamped, e.g., due to external driving. The electronic level may belong to
a neighboring quantum dot or it may be an orbital of the molecular
magnet itself. The electronic level and the molecular spin are coupled
via exchange interaction.  
We derive expressions for the noise components using the Keldysh nonequilibrium Green's functions formalism.\cite{Jauho1993,Jauho1994,JauhoBook}
The noise of charge current is contributed by both elastic processes driven by the bias voltage, and inelastic tunneling processes driven by the molecular spin precession. We observe diplike features in the shot noise due to inelastic tunneling processes and destructive quantum interference between electron transport channels involved in the spin-flip processes. The driving mechanism of the correlations of the spin-torque components in the same spatial direction involves both precession of the molecular spin and the bias voltage.
Hence, they are contributed by elastic and inelastic processes, with
the change of energy equal to one or two Larmor frequencies. The
nonzero correlations of the perpendicular spin-torque components are
driven by the molecular spin precession, with contributions of
spin-flip tunneling processes only. These components are related to
the previously obtained Gilbert damping coefficient,\cite{we1,we2}  
which characterize the Gilbert damping term of the spin-transfer
torque,\cite{Gilbert,Tserkovnyak,Ralph} at arbitrary temperature. 

The article is organized as follows. The model and theoretical
framework based on the Keldysh nonequilibrium Green's functions
formalism\cite{Jauho1993,Jauho1994,JauhoBook} are given in
Sec.~\ref{sec: model3rdpaper}. Here we derive expressions for the
noise of spin and charge currents. In Sec.~\ref{sec:
  noisechargecurrent} we investigate and analyze the properties of the
charge-current shot noise. In Sec.~\ref{sec:
  noiseofsttandspincurrent} we derive and analyze the noise
of spin-transfer torque. The conclusions are given in Sec.~\ref{sec:
  conclusions3ndpaper}.

\section{Model and theoretical framework}\label{sec: model3rdpaper}

\begin{figure}
	\includegraphics[height=3.7cm,keepaspectratio=true]{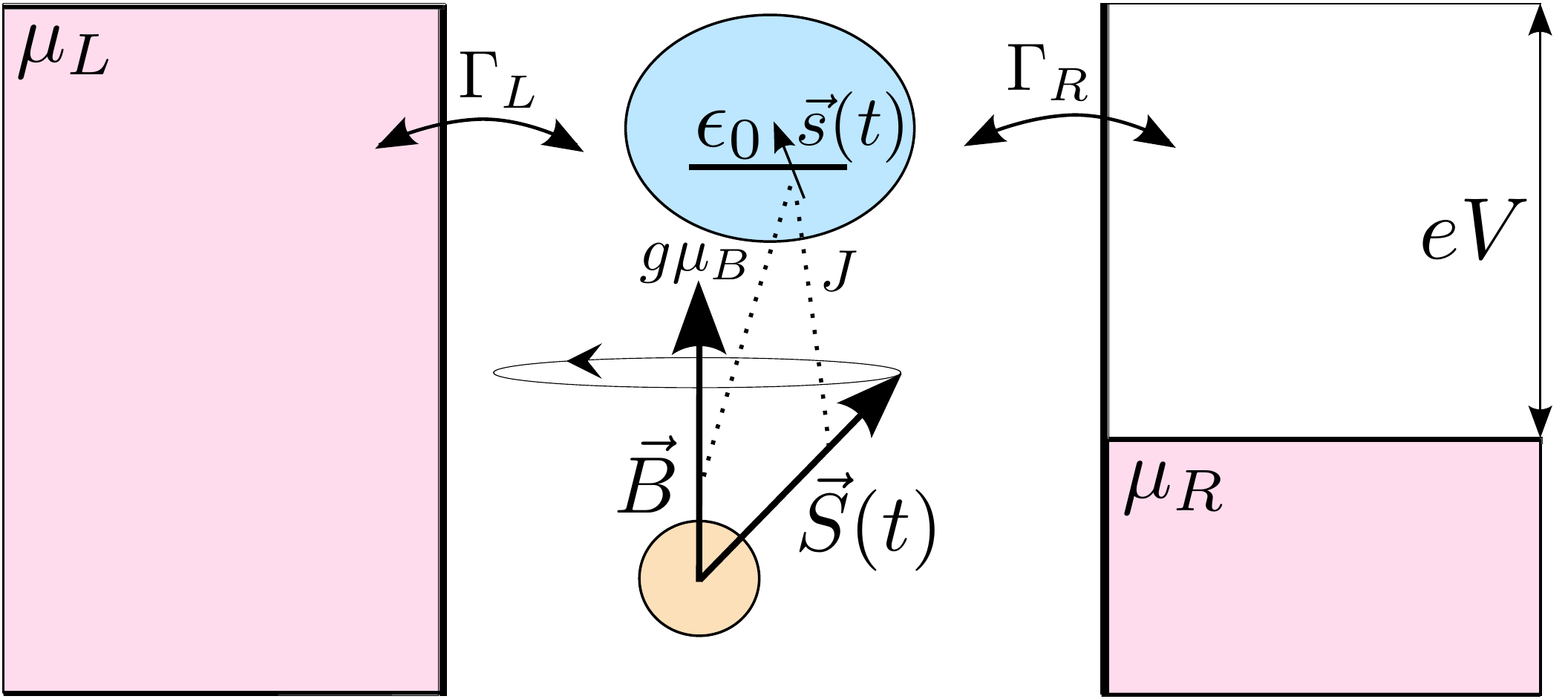}   
	\caption{(Color online) Tunneling through a single molecular level with energy $\epsilon_0$ in the presence of a precessing molecular spin $\vec{S}(t)$ in a constant magnetic field 
	$\vec{B}$, connected to two metallic leads with chemical potentials $\mu_\xi$, $\xi=L,R$. The molecular level is coupled to the spin of the molecule via exchange interaction with the coupling constant $J$. The applied dc-bias voltage $eV=\mu_{L}-\mu_{R}$, and the tunnel rates are $\Gamma_\xi$.}\label{fig: System}
\end{figure}

The junction under consideration consists of a noninteracting
single-level quantum dot in the presence of a precessing molecular
spin in a magnetic field along the $z$-axis, $\vec{B}=B\vec{e}_{z}$,
coupled to two noninteracting leads (Fig.~\ref{fig: System}).  
The junction is described by the Hamiltonian
\begin{equation}
	\hat{H}(t)=\sum_{\xi\in \{L,R\}}\hat{H}_\xi+\hat{H}_{T}+\hat{H}_{D}(t)+\hat{H}_{S},\label{eq: total}
\end{equation}
where
\begin{equation}
	\hat{H}_\xi=\sum_{k,\sigma}\epsilon_{k\xi}\hat{c}^\dagger_{k\sigma\xi} \hat{c}_{k\sigma\xi}
\end{equation}
is the Hamiltonian of contact $\xi=L,R$. The spin-(up or down) state
of the electrons is denoted by the subscript
$\sigma=\uparrow,\downarrow=1,2=\pm 1$. The tunnel coupling between
the quantum dot and the leads reads 
\begin{equation}
  \hat{H}_{T}=\sum_{k,\sigma,\xi}  [V_{k\xi}\hat{c}^\dagger_{k\sigma\xi} \hat{d}_{\sigma}+V^{\ast}_{k\xi}
  \hat{d}^\dagger_{\sigma} \hat{c}_{k\sigma\xi}],
\end{equation} 
with spin-independent matrix element $V_{k\xi}$.  
The creation (annihilation) operators of the electrons in the leads and the quantum dot are given by $ \hat{c}^\dagger_{k\sigma\xi}(\hat{c}_{k\sigma\xi})$ and $ \hat{d}^\dagger_{\sigma} (\hat{d}_{\sigma})$. The Hamiltonian of the electronic level equals 
\begin{equation}
  \hat{H}_{D}(t)=\sum_{{\sigma}}\epsilon_{0} \hat{d}^\dagger_{\sigma} \hat{d}_{\sigma}+g\mu_{B}\hat{\vec{s}}\vec B+J\hat{\vec{s}}\vec{S}(t).\label{eq: molecularorbitalHamiltonian}
\end{equation} 
The first term in Eq.~(\ref{eq: molecularorbitalHamiltonian}) is the Hamiltonian of the noninteracting single-level quantum dot with energy $\epsilon_{0}$. The second term describes the 
electronic spin in the dot,
$\hat{\vec{s}}=(\hbar/2)\sum_{\sigma\sigma'}({\vec\sigma})_{\sigma\sigma'}\hat
d^\dagger_\sigma\hat d_{\sigma'}$, in the presence of a constant
magnetic field $\vec{B}$, and the third term represents the exchange
interaction between the electronic spin and the molecular spin
$\vec{S}(t)$. The vector of the Pauli matrices is given by
$\hat{\vec\sigma}=(\hat{\sigma}_x,\hat{\sigma}_y,\hat{\sigma}_z)^T$. The
g-factor of the electron and the Bohr magneton are $g$ and
$\mu_{B}$, whereas $J$ is the  
exchange coupling constant between the electronic and molecular
spins. 
	
The last term of Eq.~(\ref{eq: total}) can be written as
\begin{equation}
  \hat{H}_{S}=g\mu_{B}{\vec S} {\vec B},
\end{equation}
and represents the energy of the
molecular spin $\vec S$ in the magnetic field $\vec B$.
We assume that $\lvert{\vec{S}}\lvert\gg\hbar$ and neglecting quantum
fluctuations treat $\vec{S}$ as a classical variable. The magnetic
field $\vec B$ generates a torque on the spin $\vec S$ that causes
the spin to precess around the field axis with Larmor frequency
$\omega_L=g\mu_{B}B/\hbar$.  
The dynamics of the molecular spin is kept constant, which can be
realized, e.g., by external rf fields\cite{Kittel} to cancel the loss of
 magnetic energy due to the interaction with the itinerant
electrons. 
Thus, the precessing spin $\vec S(t)$ pumps spin currents into the
leads, but its dynamics remains unaffected by the spin currents, i.e., 
the spin-transfer torque exerted on the molecular spin is 
compensated by the above mentioned external means. 
The undamped precessional motion of the molecular spin, supported by
the external sources, is then given by $\vec S(t)=S_{\bot}\cos
(\omega_{L} t)\vec e_x+S_{\bot}\sin (\omega_{L} t)\vec e_y+S_{z}\vec
e_z$, with $\theta$ the tilt angle between $\vec{B}$ and $\vec{S}$,
and $S_{\bot}=S\sin(\theta)$ the magnitude of the instantaneous
projection of $\vec S(t)$ onto the $xy$ plane. The component of the
molecular spin along the field axis equals $S_z=S\cos(\theta)$. 

The charge- and spin-current operators of the lead $\xi$ are given by
the Heisenberg equation \cite{JauhoBook,Jauho1994} 
\begin{equation}
\hat{I}_{\xi\nu}(t)=q_{\nu}\frac{d\hat{N}_{\xi\nu}}{dt}=q_{\nu}\frac{i}{\hbar}[\hat{H},\hat{N}_{\xi\nu}],\label{eq: commutator}
%\vspace{0.15cm}
\end{equation}
where $[\ , \ ]$ denotes the commutator, while  $\hat{N}_{L\nu}=\sum_{{k,\sigma,\sigma\prime}}
\hat{c}^\dagger_{k\sigma L}({\sigma}_\nu)_{\sigma\sigma^\prime} 
\hat{c}_{k\sigma\prime L}$ is the charge ($\nu=0$ and $q_0=-e$ ) and spin ($\nu= x,y,z$ and $q_{\nu\neq 0}=\hbar/2$) occupation number operator of the contact $\xi$. Here $\hat{\sigma}_{0}=\hat{1}$ is the identity matrix. Taking into account that only the tunneling Hamiltonian $\hat{H}_T$ generates a nonzero commutator in 
Eq.~(\ref{eq: commutator}), the current operator $\hat{I}_{\xi\nu}(t)$ can be expressed as
\begin{equation}
\hat{I}_{\xi\nu}(t)=-q_{\nu}\frac{i}{\hbar}\sum_{\sigma,\sigma'}({\sigma}_{\nu})_{\sigma\sigma'}\hat{I}_{\xi,\sigma\sigma'}(t),\label{eq: sigmacomponents}
\end{equation}
where the operator component $\hat{I}_{\xi,\sigma\sigma'}(t)$ equals
\begin{equation}
\hat{I}_{\xi,\sigma\sigma'}(t)=\sum_{k}[V_{k\xi}\hat{c}^{\dag}_{k\sigma\xi}(t)\hat{d}_{\sigma'}(t)-V^{*}_{k\xi}\hat{d}^{\dag}_{\sigma}(t)\hat{c}_{k\sigma'\xi}(t)].
\end{equation}

The nonsymmetrized noise of charge and spin current is defined as the correlation between fluctuations of currents $I_{\xi\nu}$ and $I_{\zeta\mu}$,\cite{JauhoBook,ButtikerBlanter2000}
\begin{equation}
S^{\nu\mu}_{\xi\zeta}(t,t')=\langle\delta\hat{I}_{\xi\nu}(t)\delta\hat{I}_{\zeta\mu}(t')\rangle,\label{eq: 5.4}
\end{equation}	
with $\nu=\mu=0$ for the charge-current noise. The fluctuation operator of the charge and spin current in lead $\xi$ is given by
\begin{equation}
\delta\hat{I}_{\xi\nu}(t)=\hat{I}_{\xi\nu}(t)-\langle\hat{I}_{\xi\nu}(t)\rangle.\label{eq: correlation}
\end{equation}
Using Eqs.~(\ref{eq: sigmacomponents}) and (\ref{eq: correlation}), the noise becomes
\begin{equation}
S^{\nu\mu}_{\xi\zeta}(t,t')=-\frac{q_{\nu}q_{\mu}}{\hbar^2}\sum_{\sigma\sigma'}\sum_{\lambda\eta}({\sigma}_{\nu})_{\sigma\sigma'}({\sigma}_{\mu})_{\lambda\eta}
S^{\sigma\sigma',\lambda\eta}_{\xi\zeta}(t,t'),\label{eq: 5.6}
\end{equation}
where $S^{\sigma\sigma',\lambda\eta}_{\xi\zeta}(t,t')=\langle\delta\hat{I}_{\xi,\sigma\sigma'}(t)\delta\hat{I}_{\zeta,\lambda\eta}(t')\rangle$.
The formal expression for $S^{\nu\mu}_{\xi\zeta}(t,t')$ is given by Eq.~(\ref{eq: 5.7k}) in the Appendix, where it is obtained using Eq.~(\ref{eq: 5.6}) and Eqs.~(\ref{eq: 5.7})--(\ref{eq: 5.14}).

Using Fourier transformations of the central-region Green's functions given by Eqs.~(\ref{eq: 5.8m})--(\ref{eq: 5.10m}) and self-energies in the wide-band limit, the correlations given by Eq.~(\ref{eq: 5.14}) can be further simplified. 
Some correlation functions are not just functions of time difference $t-t'$. Thus, as in Ref.~\citenum{PedersenBut}, we used Wigner representation assuming that in experiments fluctuations are measured on timescales much larger than the driving period $\mathcal{T}=2\pi/\omega_{L}$, which is the period of one molecular spin precession. The Wigner coordinates are given by $T'=(t+t')/2$ and $\tau=t-t'$, while the correlation functions are defined as
\begin{equation}
S^{\sigma\sigma',\lambda\eta}_{\xi\zeta}(\tau)=\frac{1}{\mathcal T}\int_{0}^{\mathcal T}dt\langle\delta\hat{I}_{\xi,\sigma\sigma'}(t+\tau)\delta\hat{I}_{\zeta,\lambda\eta}(t)\rangle.\label{eq: 5.15}
\end{equation}
The Fourier transform of $S^{\sigma\sigma',\lambda\eta}_{\xi\zeta}(\tau)$ is given by
\begin{equation}
S^{\sigma\sigma',\lambda\eta}_{\xi\zeta}(\Omega,\Omega')=2\pi\delta (\Omega-\Omega')S^{\sigma\sigma',\lambda\eta}_{\xi\zeta}(\Omega),
\end{equation}
where
\begin{equation}
S^{\sigma\sigma',\lambda\eta}_{\xi\zeta}(\Omega)=\int d\tau e^{i\Omega\tau}S^{\sigma\sigma',\lambda\eta}_{\xi\zeta}(\tau).\label{eq: 5.17}
\end{equation}
For the correlations which depend only on $t-t'$, the Wigner representation is identical to the standard representation.

The symmetrized noise of charge and spin currents reads \cite{JauhoBook,ButtikerBlanter2000}
\vspace*{0.05cm}
\begin{equation}
S^{\nu\mu}_{\xi\zeta S}(t,t')=\frac{1}{2}\langle\{\delta\hat{I}_{\xi\nu}(t),\delta\hat{I}_{\zeta\mu}(t')\}\rangle,\label{eq: 5.24}
\end{equation}
where $\{ , \}$ denotes the anticommutator.
According to Eqs.~(\ref{eq: 5.6}), (\ref{eq: 5.15}), (\ref{eq: 5.17}), and (\ref{eq: 5.24}), in the Wigner representation the nonsymmetrized noise spectrum reads
\begin{align}
S^{\nu\mu}_{\xi\zeta}(\Omega)&=\int d\tau e^{i\Omega\tau}S^{\nu\mu}_{\xi\zeta}(\tau)\nonumber\\
&=\int d\tau e^{i\Omega\tau}\frac{1}{\mathcal T}\int_{0}^{\mathcal T}dt\langle\delta\hat{I}_{\xi\nu}(t+\tau)\delta\hat{I}_{\zeta\mu}(t)\rangle\nonumber\\
&=-\frac{q_{\nu}q_{\mu}}{\hbar^2}\sum_{\sigma\sigma'}\sum_{\lambda\eta}({\sigma}_{\nu})_{\sigma\sigma'}({\sigma}_{\mu})_{\lambda\eta}
S^{\sigma\sigma',\lambda\eta}_{\xi\zeta}(\Omega),
\end{align}
while the symmetrized noise spectrum equals
\begin{align}
S^{\nu\mu}_{\xi\zeta S}(\Omega)&=\frac{1}{2}[S^{\nu\mu}_{\xi\zeta}(\Omega)+S^{\mu\nu}_{\zeta\xi}(-\Omega)]\nonumber\\
&=-\frac{q_{\nu}q_{\mu}}{2\hbar^2}\sum_{\sigma\sigma'}\sum_{\lambda\eta}({\sigma}_{\nu})_{\sigma\sigma'}({\sigma}_{\mu})_{\lambda\eta}
S^{\sigma\sigma',\lambda\eta}_{\xi\zeta S}(\Omega),
\end{align}
where
$S^{\sigma\sigma',\lambda\eta}_{\xi\zeta
  S}(\Omega)=[S^{\sigma\sigma',\lambda\eta}_{\xi\zeta}(\Omega)+S^{\lambda\eta,\sigma\sigma'}_{\zeta\xi}(-\Omega)]/2$. 
The experimentally most easily accessible quantity is the  zero-frequency noise
power. 

\section{Shot noise of charge current}\label{sec: noisechargecurrent}

\begin{figure*}[t]
	\includegraphics[height=5.4cm,keepaspectratio=true]{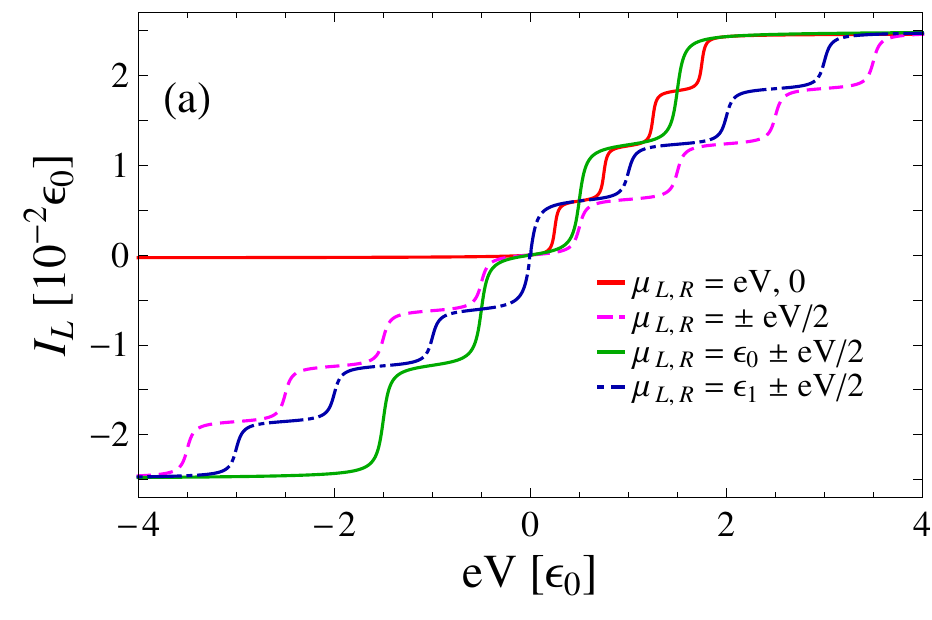}\,\,\,\,  
	\includegraphics[height=5.4cm,keepaspectratio=true]{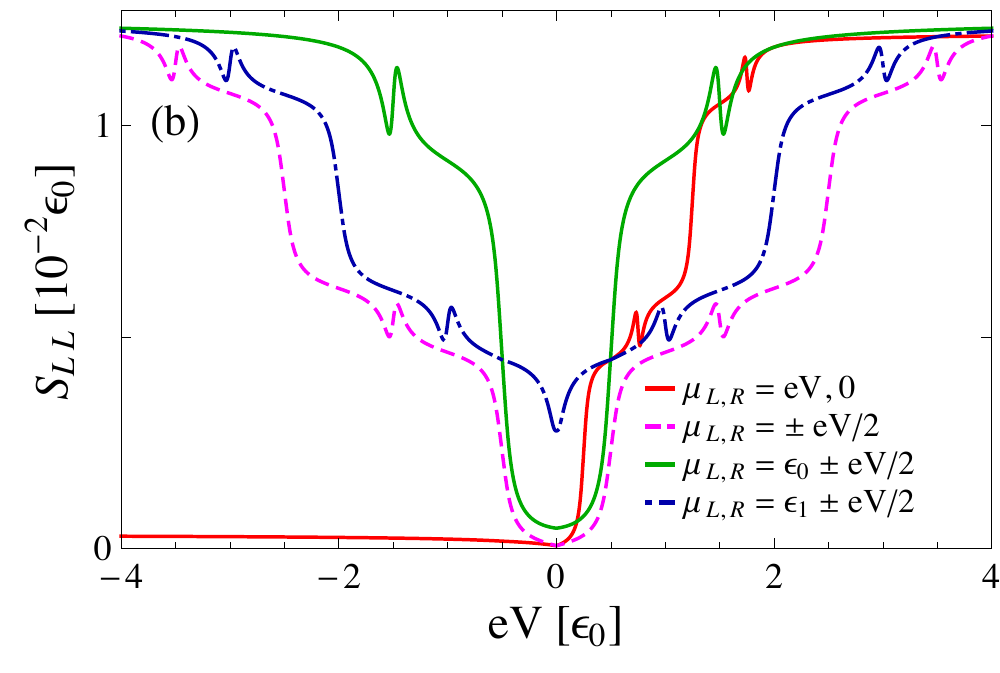}    
	\caption{(Color online) (a) Charge current $I_{L}$ and (b) auto-correlation shot noise $S_{LL}$ as functions of bias-voltage $eV$. All plots are obtained at zero temperature, with $\vec{B}=B\vec{e}_{z}$. The other parameters are $\Gamma_{L}=\Gamma_{R}=\Gamma/2$, $\Gamma=0.05\,\epsilon_{0}$, $\omega_{L}=0.5\,\epsilon_{0}$, $J=0.01\,\epsilon_{0}$, $S=100$, and $\theta=\pi/2$. The molecular quasienergy levels are located at $\epsilon_{1}=0.25\,\epsilon_0$, $\epsilon_{2}=0.75\,\epsilon_0$,
		$\epsilon_{3}=1.25\,\epsilon_0$, and $\epsilon_{4}=1.75\,\epsilon_0$.
	}\label{fig: noisecharge}
\end{figure*}
For the charge-current noise, it is convenient to drop the superscripts $\nu=\mu=0$. The charge-current noise spectrum can be obtained as
\cite{Sauret2004}
%\begin{equation}
%	%S_{\xi\zeta}(\Omega)=-\frac{e^{2}}{\hbar^2}\sum_{\sigma\sigma'}S^{\si %gma\sigma,\sigma'\sigma'}_{\xi\zeta}(\Omega).
%\end{equation}
\begin{equation}
	S_{\xi\zeta}(\Omega)=-\frac{e^{2}}{\hbar^2}[S^{11,11}_{\xi\zeta}+S^{11,22}_{\xi\zeta}+S^{22,11}_{\xi\zeta}+S^{22,22}_{\xi\zeta}](\Omega).
\end{equation}
In this section, we analyze the zero-frequency %$\Omega=0$ 
noise power of the charge current $S_{\xi\zeta}=S_{\xi\zeta}(0)$ at
zero temperature. Taking into account that thermal noise disappears at
zero temperature, the only contribution to the charge-current noise
comes from the shot noise.  
The tunnel couplings between the molecular orbital and the leads,
$\Gamma_\xi(\epsilon)=2\pi\sum_{k}\lvert V_{k\xi}\rvert^{2}\delta(\epsilon-\epsilon_{k\xi})$, are considered  symmetric and in the wide-band limit 
 $\Gamma_{L}=\Gamma_{R}=\Gamma/2$. 

The average charge current from lead $\xi$ can be expressed as
\begin{align}
I_{\xi}=&\frac{e\Gamma_{\xi}\Gamma_{\zeta}}{\hbar}\int\frac{d\epsilon}{2\pi}[f_{\xi}(\epsilon)-f_{\zeta}(\epsilon)]\nonumber\\
&\times\sum_{\substack{\sigma\sigma' \\ \sigma\neq\sigma'}}
\frac{\lvert G^{0r}_{\sigma\sigma}(\epsilon)\rvert^{2}[1+\gamma^{2}\lvert G^{0r}_{\sigma'\sigma'}(\epsilon+\sigma'\omega_{L})\rvert^2]}{\lvert 1-\gamma^{2} G^{0r}_{\sigma\sigma}(\epsilon)G^{0r}_{\sigma'\sigma'}(\epsilon+\sigma'\omega_{L})\rvert^2},
\end{align}
where $\xi\neq\zeta$, while ${G}^{0r}_{\sigma\sigma}(\epsilon)$ are matrix elements of 
  $\hat{G}^{0r}(\epsilon)=[\epsilon-\epsilon_{0}+i\sum_{\xi}\Gamma_{\xi}/2-\hat{\sigma}_{z}(g\mu_{B}B+JS_{z})/2]^{-1}$.\cite{Guo,Bode} 
In the above expression, $f_{\xi}(\epsilon)=[e^{(\epsilon-\mu_{\xi})/k_{B}T}+1]^{-1}$ is the Fermi-Dirac distribution of the electrons in lead $\xi$, with $k_{B}$ the Boltzmann constant and $T$ the temperature.
The conservation of the charge current implies that 
$S_{LL}(0)+S_{LR}(0)=0$. Thus, it is sufficient to study only one correlation function.

Tunning the parameters in the system such as the bias voltage $eV=\mu_{L}-\mu_{R}$ (where $\mu_{L}$ and $\mu_{R}$ are the chemical potentials of the leads), $\vec{B}$, and the tilt angle $\theta$, the shot noise can be controlled and minimized. The shot noise in the small precession frequency limit $\omega_{L}\ll k_{B}T$ is in agreement with Ref.~\citenum{WangWang2004} for $eV=0$.

In Fig.~\ref{fig: noisecharge}(a) we present the average charge
current as a staircase function of bias voltage, where the bias is
varied in four different ways. In the presence of the external
magnetic field and the precessing molecular spin, the initially
degenerate electronic level with energy $\epsilon_0$ results in four
nondegenerate transport channels, which has an important influence on
the noise. Each step corresponds to a new available transport
channel. The transport channels are located  
at the Floquet quasienergies\cite{we2}
$\epsilon_{1}=\epsilon_{0}-(\omega_{L}/2)-(JS/2)$,
$\epsilon_{2}=\epsilon_{0}+(\omega_{L}/2)-(JS/2)$,
$\epsilon_{3}=\epsilon_{0}-(\omega_{L}/2)+(JS/2)$, and
$\epsilon_{4}=\epsilon_{0}+(\omega_{L}/2)+(JS/2)$, 
which are calculated using the Floquet
theorem.\cite{Floquet1,Floquet2,Floquet3,Floquet4,Floquet5} 

The correlated current fluctuations give nonzero noise power, which is
presented in Fig.~\ref{fig: noisecharge}(b). The noise power shows the
molecular quasienergy spectrum, and each step or diplike feature in
the noise denotes the energy of a new available transport channel. The
noise has two steps and two diplike features that correspond to
these resonances. Charge current and noise power are saturated for
large bias voltages. If the Fermi levels of the leads lie below the
resonances, the shot noise approaches zero for $eV\rightarrow 0$ [red
and dashed pink lines in Fig.~\ref{fig: noisecharge}(b)]. This is due
to the fact that a small number of electron states can participate in
transport inside this small bias window and both current and noise are
close to 0. If the bias voltage is varied with respect to the resonant
energy $\epsilon_{1}$ such that $\mu_{L,R}=\epsilon_{1}\pm eV/2$
[dot-dashed blue line in Fig.~\ref{fig: noisecharge}(b)], or 
with respect to $\epsilon_0$ such that $\mu_{L,R}=\epsilon_{0}\pm
eV/2$ [green line in Fig.~\ref{fig: noisecharge}(b)],  
we observe a valley at zero bias $eV=0$, which corresponds to
$\mu_{L}=\mu_{R}=\epsilon_{1}$ in the first case, and  
nonzero noise in the second case. 
For $eV=0$, the charge current is zero, but the precession-assisted
inelastic processes involving the absorption of an energy quantum
$\omega_{L}$ give rise to the noise here.  

\begin{figure}
	\includegraphics[height=5.4cm,keepaspectratio=true]{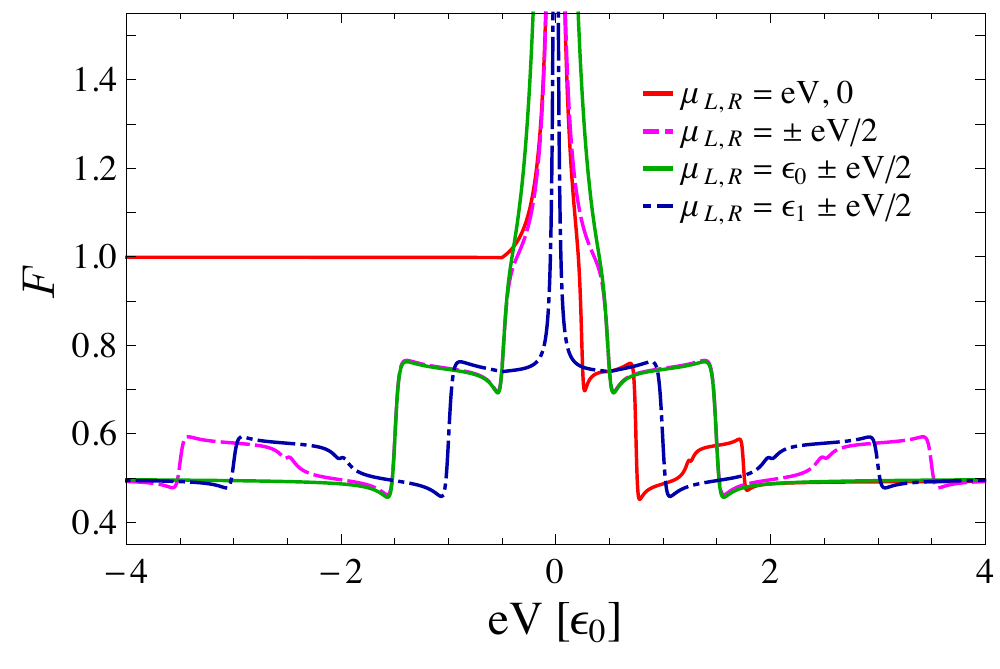}   
	\caption{(Color online) Fano factor $F$ as a function of bias-voltage $eV$. All plots are obtained at zero temperature, with $\vec{B}=B\vec{e}_{z}$. The other parameters are set to $\Gamma=\nobreak0.05\,\epsilon_{0}$, $\Gamma_{L}=\Gamma_{R}=\Gamma/2$, $\omega_{L}=0.5\,\epsilon_{0}$, $J=0.01\,\epsilon_{0}$, $S=100$, and $\theta=\pi/2$. The positions of the molecular quasienergy levels are $\epsilon_{1}=0.25\,\epsilon_0$, $\epsilon_{2}=0.75\,\epsilon_0$,
		$\epsilon_{3}=1.25\,\epsilon_0$, and $\epsilon_{4}=1.75\,\epsilon_0$.}\label{fig: Fano}
\end{figure}

At small bias voltage, 
the Fano factor $F=S_{LL}/e\lvert I_{L} \rvert$ is
inversely proportional to $eV$ and hence diverges as $eV\rightarrow
0$, indicating that the noise is super-Poissonian, as depicted in
Fig.~\ref{fig: Fano}. Due to absorption (emission) processes \cite{Floquet4} and quantum
interference effects, the Fano factor is a deformed steplike function,
where each step corresponds to a resonance.  
As the bias voltage is increased, the noise is enhanced since the
number of the correlated electron pairs increases with the increase of
the Fermi level. For larger bias, due to the absorption and emission
of an energy quantum $\omega_{L}$, electrons can jump to a level with
higher energy or lower level during the transport, and the Fano factor
$F<1$ indicates the sub-Poissonian noise. 
Around the resonances $\mu_{L,R}=\epsilon_{i}$, $i=1,2,3,4$, the
probability of transmission is very high, resulting in the small Fano
factor. Elastic tunneling contributes to the sub-Poissonian Fano
factor around the resonances and competes with the spin-flip events caused
by the molecular spin precession. However, if the resonant quasienergy
levels are much higher than the Fermi energy of the leads, the
probability of transmission is very low and the Fano factor is close
to 1, as shown in Fig.~\ref{fig: Fano} (red line). This means that the
stochastic processes are uncorrelated. If the two levels connected
with the inelastic photon emission (absorption) tunnel processes, or
all four levels, lie between the Fermi levels of the leads, the Fano
factor approaches 1/2, which is in agreement with
Ref.~\citenum{Thielmann}. For $eV=\epsilon_{3}$ [see  
Fig.~\ref{fig: Fano} (red line)] a spin-down electron can tunnel
elastically, or inelastically in a spin-flip process, leading to the
increase of the Fano factor.   
Spin-flip processes increase the electron traveling time, leading to
sub-Poissonian noise. Similarly, the Pauli exclusion principle is known
to  lead to sub-Poissonian noise, since it prevents the double occupancy of a
level. 
\begin{figure}
	\includegraphics[height=5.4cm,keepaspectratio=true]{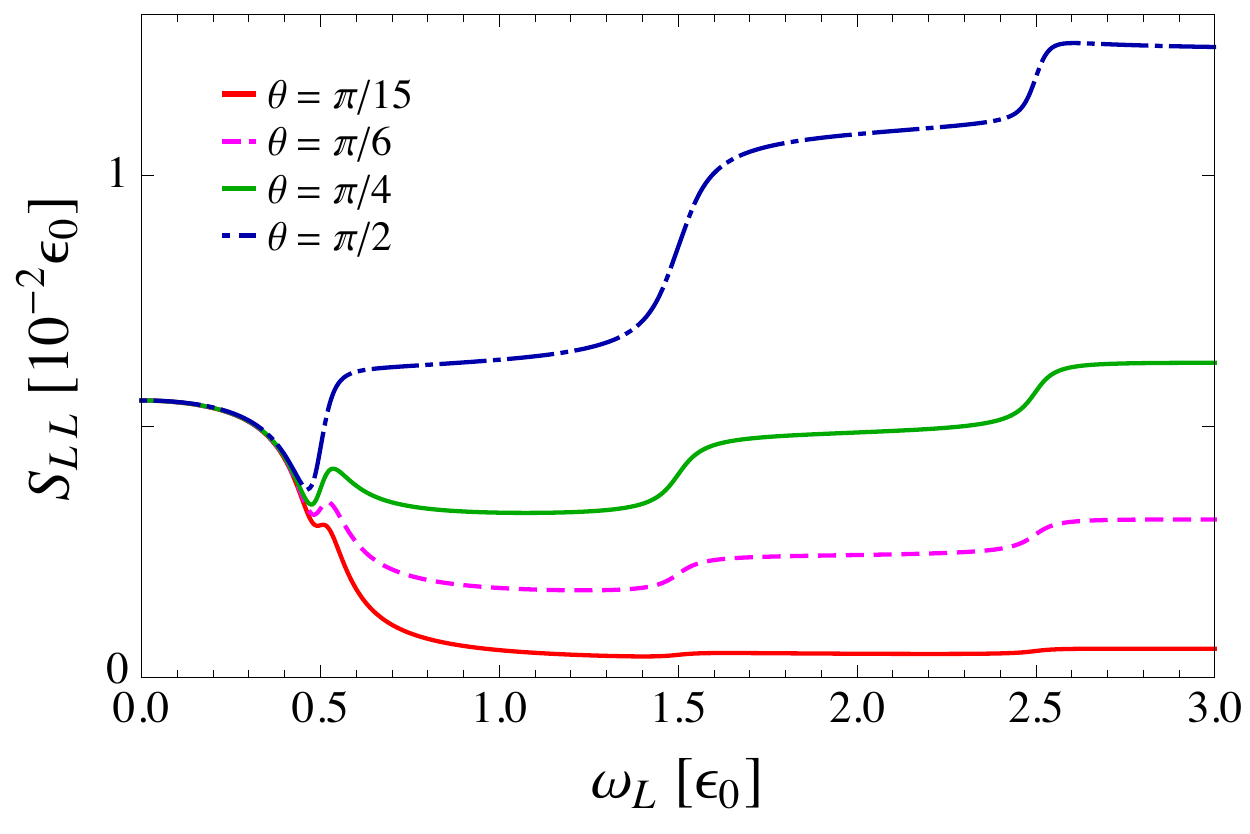}
	\caption{(Color online) Shot noise of charge current $S_{LL}$ as a function of the Larmor frequency $\omega_L$ for different tilt angles $\theta$, with $\vec{B}=B\vec{e}_{z}$, at zero temperature. The other parameters are $\Gamma=0.05\,\epsilon_{0}$, $\Gamma_{L}=\Gamma_{R}=\Gamma/2$, $\mu_{L}=0.75\,\epsilon_{0}$, $\mu_{R}=0.25\,\epsilon_{0}$, $J=0.01\,\epsilon_{0}$, and $S=100$. For $\omega_{L}=\mu_{L}-\mu_{R}$, we observe a dip due to destructive quantum interference.}\label{fig: noiseteza10}
\end{figure}
\begin{figure}
	\includegraphics[height=5.4cm,keepaspectratio=true]{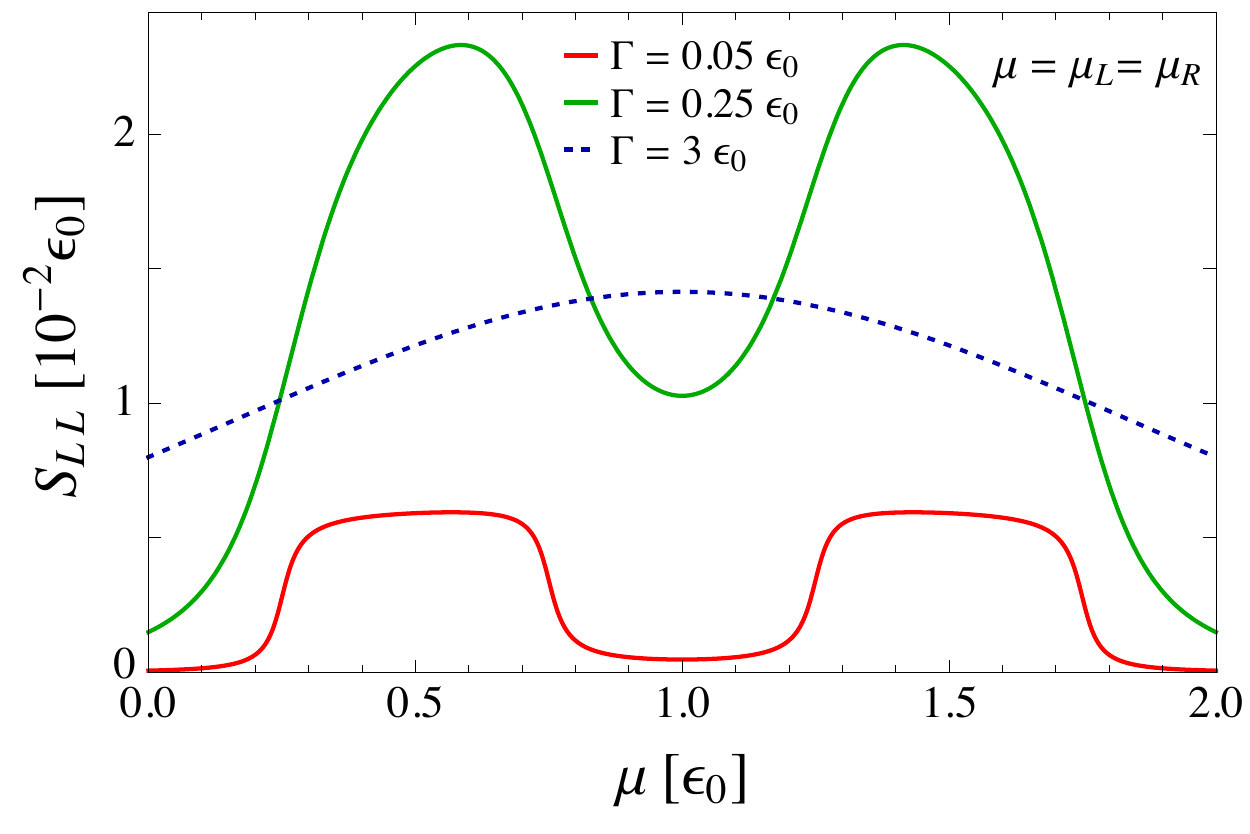}
	\caption{(Color online) Shot noise of charge current $S_{LL}$ as a function of the chemical potential of the leads $\mu=\mu_{L}=\mu_{R}$, with $\vec{B}=B\vec{e}_{z}$, for three different couplings $\Gamma$, where $\Gamma_{L}=\Gamma_{R}=\Gamma/2$, at zero temperature. The other parameters are $\omega_{L}=0.5\,\epsilon_{0}$, $J=0.01\,\epsilon_{0}$, $S=100$, and $\theta=\pi/2$. The molecular quasienergy levels are positioned at $\epsilon_{1}=0.25\,\epsilon_0$, $\epsilon_{2}=0.75\,\epsilon_0$,
		$\epsilon_{3}=1.25\,\epsilon_0$, and $\epsilon_{4}=1.75\,\epsilon_0$.}\label{fig: noisezerobias}
\end{figure}

The precessing molecular spin induces quantum interference 
between the transport channels connected with spin-flip events and the change of energy by one energy quantum $\omega_L$, i.e., between levels with energies $\epsilon_{1}$ and $\epsilon_{2}=\epsilon_{1}+\omega_{L}$, or $\epsilon_{3}$ and $\epsilon_{4}=\epsilon_{3}+\omega_{L}$. 
The destructive quantum-interference effects manifest themselves
in the form of diplike features in Fig.~\ref{fig: noisecharge}(b).
When one or both pairs of the levels connected with spin-flip events
enter the bias-voltage window, then an electron from the left lead can
tunnel through both levels via elastic or inelastic spin-flip
processes.  
Different tunneling pathways ending in the final state with the same
energy destructively interfere, similarly as in the Fano
effect.\cite{Fano1961}  
Namely, the state with lower energy $\epsilon_{1}$ (or $\epsilon_3$)
mimics the discrete state in the Fano effect. An electron tunnels into
the state $\epsilon_1$ (or $\epsilon_3$), undergoes a spin flip and
absorbs an energy quantum $\omega_L$. The other state with energy
$\epsilon_2$ (or $\epsilon_4$) is an analog of the continuum in the
Fano effect, and the electron tunnels elastically through this
level. These two tunneling processes (one elastic and the other
inelastic) interfere, leading to a diplike feature in the noise
power. 
If we vary, for instance, the bias-voltage as $eV=\mu_L$, where
$\mu_{R}=0$ [Fig.~\ref{fig: noisecharge}(b), red line], we observe
diplike features for $eV=\epsilon_2$ and $eV=\epsilon_4$.  

The destructive interference effect is also presented in
Fig.~\ref{fig: noiseteza10}, where noise power $S_{LL}$ is depicted as
a function of $\omega_L$. 
Here, we observe a dip due to the quantum-interference effect around
$\omega_L=0.5\,\epsilon_0$, which corresponds to $\mu_{L}=\epsilon_2$
and $\mu_{R}=\epsilon_1$.  
The other two steps in Fig.~\ref{fig: noiseteza10} occur when the
Fermi energy of the right or left lead is in resonance with one of the
quasienergy levels. The magnitude of the precessing component of the
molecular spin, which induces spin-flip processes between molecular
quasienergy levels, equals $JS\sin(\theta)/2$. Therefore, the dip
increases with the increase of the tilt angle $\theta$, and it is maximal
and distinct for $\theta=\pi/2$. 

Finally, in Fig.~\ref{fig: noisezerobias} we plotted the noise power
of charge current $S_{LL}$ as a function of $\mu=\mu_{L}=\mu_{R}$ at
zero temperature. It shows a nonmonotonic dependence on the tunneling
rates $\Gamma$. For small $\Gamma$ (Fig.~\ref{fig: noisezerobias}, red
line) the noise is increased if $\mu$ is positioned between levels
connected with spin-flip events, and is contributed only by absorption
processes of an energy quantum $\omega_L$ as we vary the chemical
potentials. For larger $\Gamma$ (Fig.~\ref{fig: noisezerobias}, green
line), the charge-current noise is increased since levels broaden and
overlap, and more electrons can tunnel. With further increase of
$\Gamma$ (Fig.~\ref{fig: noisezerobias}, dotted blue line) the noise
starts to decrease, and it is finally suppressed for
$\Gamma\gg\omega_L$ since a current-carrying electron sees the
molecular spin as nearly static in this case, leading to a reduction
of the inelastic spin-flip processes. 

\section{Shot noise of spin current and spin-transfer torque}\label{sec: noiseofsttandspincurrent}

In this section we present the spin-current noise spectrum components
and relations between them. Later we introduce the noise of
spin-transfer torque and we investigate the zero-frequency spin-torque
shot noise at zero temperature. The components of the nonsymmetrized
spin-current noise spectrum read  
\begin{align}
S^{xx}_{\xi\zeta}(\Omega)&=-\frac{1}{4}[S^{12,21}_{\xi\zeta}+S^{21,12}_{\xi\zeta}](\Omega),\\
S^{xy}_{\xi\zeta}(\Omega)&=-\frac{i}{4}[S^{12,21}_{\xi\zeta}-S^{21,12}_{\xi\zeta}](\Omega),\\
S^{zz}_{\xi\zeta}(\Omega)&=-\frac{1}{4}[S^{11,11}_{\xi\zeta}-S^{11,22}_{\xi\zeta}-S^{22,11}_{\xi\zeta}+S^{22,22}_{\xi\zeta}](\Omega),\label{eq: treca24maj}	
\end{align}
where Eq.~({\ref{eq: treca24maj}}) denotes the noise of the $z$ component of the spin current.\cite{Sauret2004,WangWang2004} Since the polarization of the spin current precesses in the $xy$ plane, the remaining components of the spin-current noise spectrum satisfy the following
relations:
\begin{align}
%	\vspace*{-1cm}
S^{yy}_{\xi\zeta}(\Omega)&=S^{xx }_{\xi\zeta}(\Omega),\label{eq: 5.35ok}\\	
S^{yx}_{\xi\zeta}(\Omega)&=-S^{xy}_{\xi\zeta}(\Omega),\label{eq: 5.31}\\
S^{xz}_{\xi\zeta}(\Omega)&=S^{zx}_{\xi\zeta}(\Omega)=S^{yz}_{\xi\zeta}(\Omega)=S^{zy}_{\xi\zeta}(\Omega)=0.	
\end{align}	

Taking into account that the spin current is not a conserved quantity,
it is important to notice that the complete information from the noise
spectrum can be obtained by studying both the autocorrelation noise
spectrum $S^{jk}_{\xi\xi}(\Omega)$ and cross-correlation noise
spectrum $S^{jk}_{\xi\zeta}(\Omega)$, $\zeta\neq\xi$. Therefore, it is
more convenient to investigate the spin-torque noise spectrum, where
both autocorrelation and cross-correlation noise components of spin
currents are included. The spin-transfer torque operator can be
defined as 
\begin{equation}
\hat{T}_{j}=-(\hat{I}_{Lj}+\hat{I}_{Rj}),\qquad\qquad j=x,y,z;
\end{equation}
while its fluctuation reads
\begin{equation}
\delta\hat{T}_{j}(t)=-[\delta\hat{I}_{Lj}(t)+\delta\hat{I}_{Rj}(t)].
\end{equation}
Accordingly, the nonsymmetrized and symmetrized spin-torque noise can be obtained using the spin-current noise components as
\begin{align}
S^{jk}_{T}(t,t')&=\langle\delta\hat{T}_{j}(t)\delta\hat{T}_{k}(t')\rangle\nonumber\\
&=\sum_{\xi\zeta}S^{jk}_{\xi\zeta}(t,t'),\qquad\qquad j,k=x,y,z;\\
S^{jk}_{TS}(t,t')&=\frac{1}{2}[S^{jk}_{T}(t,t')+S^{kj}_{T}(t',t)],
\end{align}
with the corresponding noise spectrums given by
\begin{align}
S^{jk}_{T}(\Omega)&=\sum_{\xi\zeta}S^{jk}_{\xi\zeta}(\Omega),\label{eq: 5.43ok}\\
S^{jk}_{TS}(\Omega)&=\sum_{\xi\zeta}S^{jk}_{\xi\zeta S}(\Omega).\label{eq: 5.39}
\end{align}
According to Eqs.\,(\ref{eq: 5.35ok}),\,(\ref{eq: 5.31}), and
(\ref{eq: 5.43ok}), $S^{xx}_{T}(\Omega)=S^{yy}_{T}(\Omega)$ and
$S^{yx}_{T}(\Omega)=-S^{xy}_{T}(\Omega)$. 
 
In the remainder of the section we investigate the zero-frequency spin-torque shot noise
$S^{jk}_{T}=S^{jk}_{T}(0)$ at zero temperature, where $S^{xx}_{T}(0)=S^{xx}_{TS}(0)$, $S^{yy}_{T}(0)=S^{yy}_{TS}(0)$, $S^{zz}_{T}(0)=S^{zz}_{TS}(0)$, while $S^{xy}_{T}(0)$ is a complex imaginary function, and $S^{xy}_{TS}(0)=0$ according to Eqs.~(\ref{eq: 5.31}) and (\ref{eq: 5.39}). Since $S^{xx}_{T}(0)=S^{yy}_{T}(0)$, all results and discussions related to $S^{xx}_{T}(0)$ also refer to $S^{yy}_{T}(0)$.

Spin currents $I_{\xi x}$ and $I_{\xi y}$ are periodic functions of
time, with period $\mathcal{T}=2\pi/\omega_{L}$, while $I_{\xi z}$ is
time-independent. It has already been demonstrated that spin-flip
processes contribute to the noise of spin current.\cite{WangWang2004}
The presence of the precessing molecular spin affects the spin-current
noise. Since the number of particles with different spins changes due
to spin-flip processes, additional spin-current fluctuations are
generated. Currents with the same and with different spin orientations
are correlated during transport.  
%Spin-flip processes induce correlations of currents with opposite spins. 
%Spin currents are auto-correlated in the sense that two spin-currents polarized along the same direction are correlated, while $I_{\xi x}$ and $I_{\zeta y}$ are also correlated between themselves. 
Due to the precessional motion of the molecular spin, inelastic spin
currents with spin-flip events induce noise of spin currents and
spin-torque noise, which can be nonzero even for $eV=0$.  
%The correlations between the torques in the same direction induce spin-torque noise, which is nonzero even for $eV=0$, due to the molecular spin precession. 
%Nonequilibrium precession-assisted noise of spin-transfer torque at zero temperature is the subject of our investigation here. 
The noise component $S^{xy}_{T}$ is induced by the molecular spin
precession and vanishes for a static molecular spin. The noises of
spin currents and spin-transfer torque are driven by the bias voltage
and by the molecular spin precession. Hence, in the case when both the
molecular spin is static (absence of inelastic spin-flip processes)
and $eV=0$ (no contribution of elastic tunneling processes), they
are all equal to zero.  
The noise of spin-transfer torque can be modified by adjusting system
parameters such as the bias voltage $eV$, the magnetic field
$\vec{B}$, or the tilt angle $\theta$.  
\begin{figure}
	\includegraphics[height=5.4cm,keepaspectratio=true]{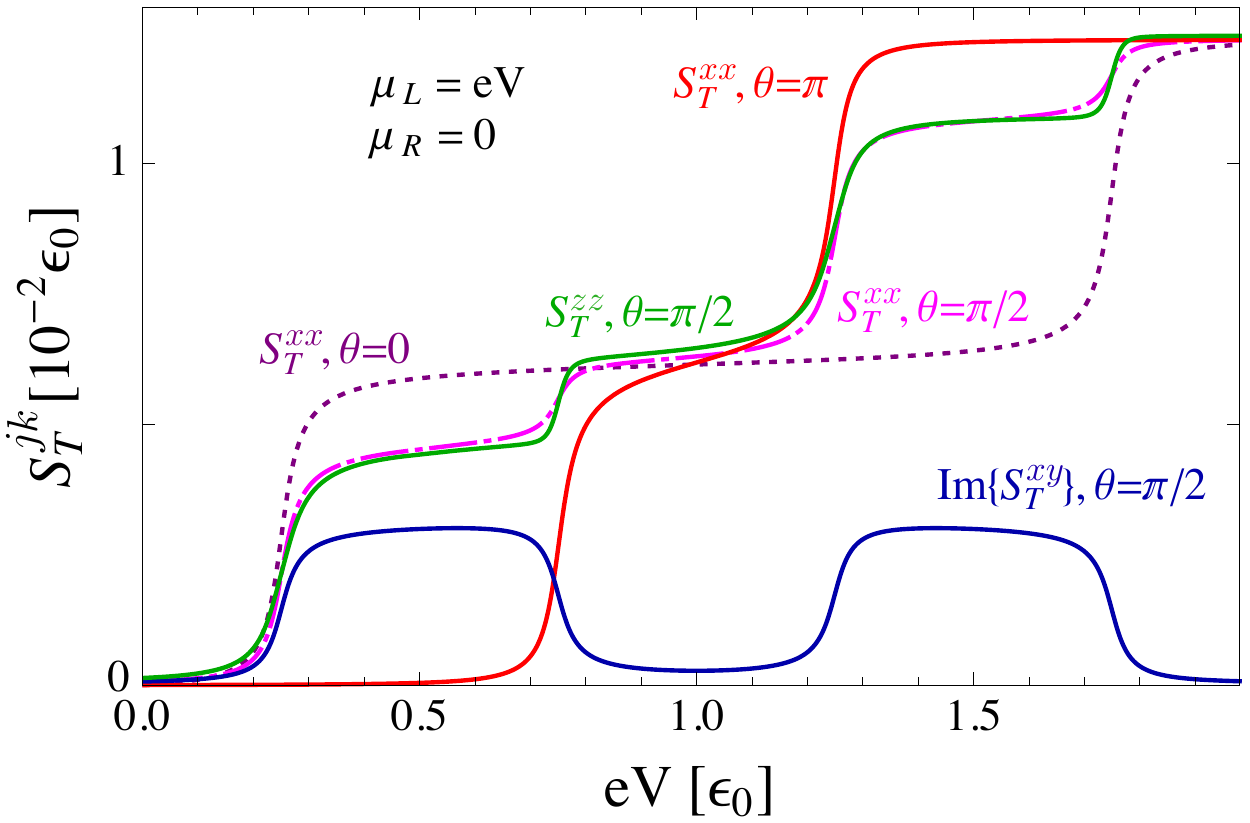}
	\caption{(Color online) Spin-torque shot-noise components $S^{jk}_T$ as functions of the bias voltage $eV$ for $\mu_{R}=0$, $\mu_{L}=eV$. All plots are obtained at zero temperature, with $\vec B=B\vec e_{z}$, and $\Gamma_{L}=\Gamma_{R}=\Gamma/2$, for $\Gamma=0.05\,\epsilon_{0}$. 
		The other parameters are set to $\omega_{L}=0.5\,\epsilon_{0}$, $J=0.01\,\epsilon_{0}$, and $S=100$. The molecular quasienergy levels lie at $\epsilon_{1}=0.25\,\epsilon_0$, $\epsilon_{2}=0.75\,\epsilon_0$,
		$\epsilon_{3}=1.25\,\epsilon_0$, and $\epsilon_{4}=1.75\,\epsilon_0$.}\label{fig: spintorquebias}
\end{figure}

In Fig.~\ref{fig: spintorquebias} we present the zero-frequency
spin-torque noise components $S^{xx}_{T}=S^{yy}_{T}$, ${\rm
  Im}\{S^{xy}_{T}\}$, and $S^{zz}_{T}$ as functions of the bias
voltage $eV=\mu_{L}-\mu_{R}$, for $\mu_{R}=0$ and different tilt
angles $\theta$ between $\vec{B}$ and $\vec{S}$ at zero
temperature. They give information on available transport channels and
inelastic spin-flip processes. The magnitude of the torque noise at
resonance energies $\epsilon_i$, $i=1,2,3,4$, is determined by
$\theta$.  In cases $\theta=0$ and $\theta=\pi$, there are only two
transport channels of opposite spins determined by the resulting
Zeeman field $B\pm JS/g\mu_{B}$. The component $S^{xx}_{T}$ shows two
steps with equal heights located at these resonances, where the only
contribution to the spin-torque noise comes from elastic tunneling
events 
(dotted purple and red lines in Fig.~\ref{fig: spintorquebias}). 
For $\theta=\pi/2$, the elastic tunneling contributes with four steps with equal heights located at resonances $\epsilon_{i}$, but due to the contributions of the inelastic precession-assisted processes between quasienergy levels $\epsilon_1$($\epsilon_3$) and $\epsilon_2$($\epsilon_4$), the heights of the steps in $S^{xx}_{T}$ are not equal anymore (dot-dashed pink line in Fig.~\ref{fig: spintorquebias}). 
Here, we observed that the contribution of the inelastic tunneling
processes to $S^{xx}_{T}$, involving absorption of an energy quantum
$\omega_L$ and a spin-flip, shows steps at spin-down quasienergy
levels $\epsilon_{1}$ and $\epsilon_{3}$, while it is constant between
and after the bias has passed these levels. 
%We observed that the contribution of the one quantum $\omega_L$ absorption tunneling processes to $S^{xx}_{T}$ shows steps at spin-down levels $\epsilon_{1}$ and $\epsilon_{3}$, while it is constant between and after the bias has passed these levels. 
The component $S^{zz}_{T}$ shows similar behavior (green line in
Fig.~\ref{fig: spintorquebias}). As in the case of the
inelastic tunneling involving the absorption of one energy quantum
$\omega_L$, in $S^{xx}_{T}=S^{yy}_{T}$ we observed inelastic spin-flip
processes involving the absorption of two energy quanta $2\omega_L$
in the form of steps at spin-down levels $\epsilon_{1}$,
$\epsilon_{3}$, $\epsilon_{2}-2\omega_{L}$, and
$\epsilon_{4}-2\omega_{L}$, which have negligible contribution
compared to the other terms. These processes are a result of
correlations of two oscillating spin-currents.  
For large bias voltage, the spin-torque noise components $S^{xx}_{T}$
and $S^{zz}_{T}$  saturate. 

The behavior of the component ${\rm Im}\{S^{xy}_{T}\}$ is completely
different in nature. It is contributed only by one energy quantum
$\omega_L$ absorption (emission) spin-flip processes. Interestingly,
we obtained the following relation between the Gilbert damping
parameter $\alpha$,\cite{we1,we2} and ${\rm Im}\{S^{xy}_{T}\}$ at
arbitrary temperature  
%at zero temperature
\begin{equation}
{\rm Im}\{S^{xy}_{T}\}=\frac{\omega_{L}S\sin^{2}(\theta)}{2}\alpha.
\end{equation}
Hence, the component ${\rm Im}\{S^{xy}_{T}\}$ is increased for Fermi
levels of the leads positioned in the regions where inelastic
tunneling processes occur (blue line in Fig.~\ref{fig:
  spintorquebias}). 
\begin{figure}
	\includegraphics[height=5.4cm,keepaspectratio=true]{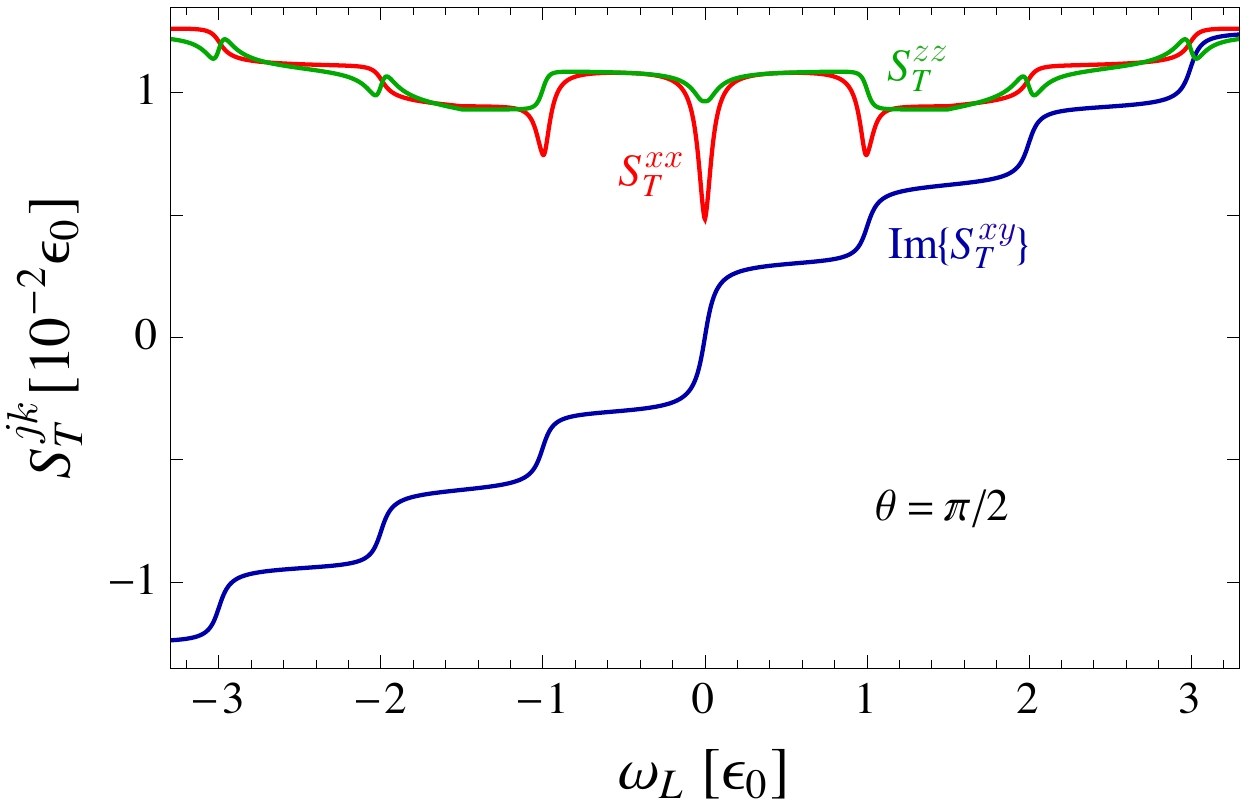}
	\caption{(Color online) Spin-torque shot-noise components $S^{jk}_T$ as functions of the Larmor frequency $\omega_{L}$ for $\theta=\pi/2$, $\mu_{R}=0$, and $\mu_{L}=1.5\,\epsilon_{0}$. All plots are obtained for $\vec B=B\vec e_{z}$ at zero temperature. The other parameters are $\Gamma_{L}=\Gamma_{R}=\Gamma/2$, $\Gamma=0.05\,\epsilon_{0}$, $J=0.01\,\epsilon_{0}$, and $S=100$.}\label{fig: NoiseLarmor}
\end{figure}

The spin-torque noise is influenced by the magnetic field $\vec{B}$
since it determines the spin-up and spin-down molecular quasienergy
levels. The dependence of $S^{xx}_{T}$, ${\rm Im}\{S^{xy}_{T}\}$, and
$S^{zz}_{T}$ on the Larmor frequency $\omega_{L}$ is depicted in
Fig.~\ref{fig: NoiseLarmor}. The steps, dips, or peaks in the plots are
located at resonant tunneling frequencies $\omega_{L}=\pm\lvert
2\mu_{L,R}-2\epsilon_{0}\pm JS\rvert$. For $\omega_{L}=0$ there are
only two transport channels, one at energy $\epsilon_{0}+JS/2$, which
is equal to the Fermi energy of the left lead, and the other at
$\epsilon_{0}-JS/2$ located between $\mu_{L}$ and $\mu_{R}$. The
contributions of the elastic spin transport processes through these
levels result in dips in the components $S^{xx}_{T}$ and $S^{zz}_{T}$,
while ${\rm Im}\{S^{xy}_{T}\}=0$.  
\begin{figure}
	\includegraphics[height=5.4cm,keepaspectratio=true]{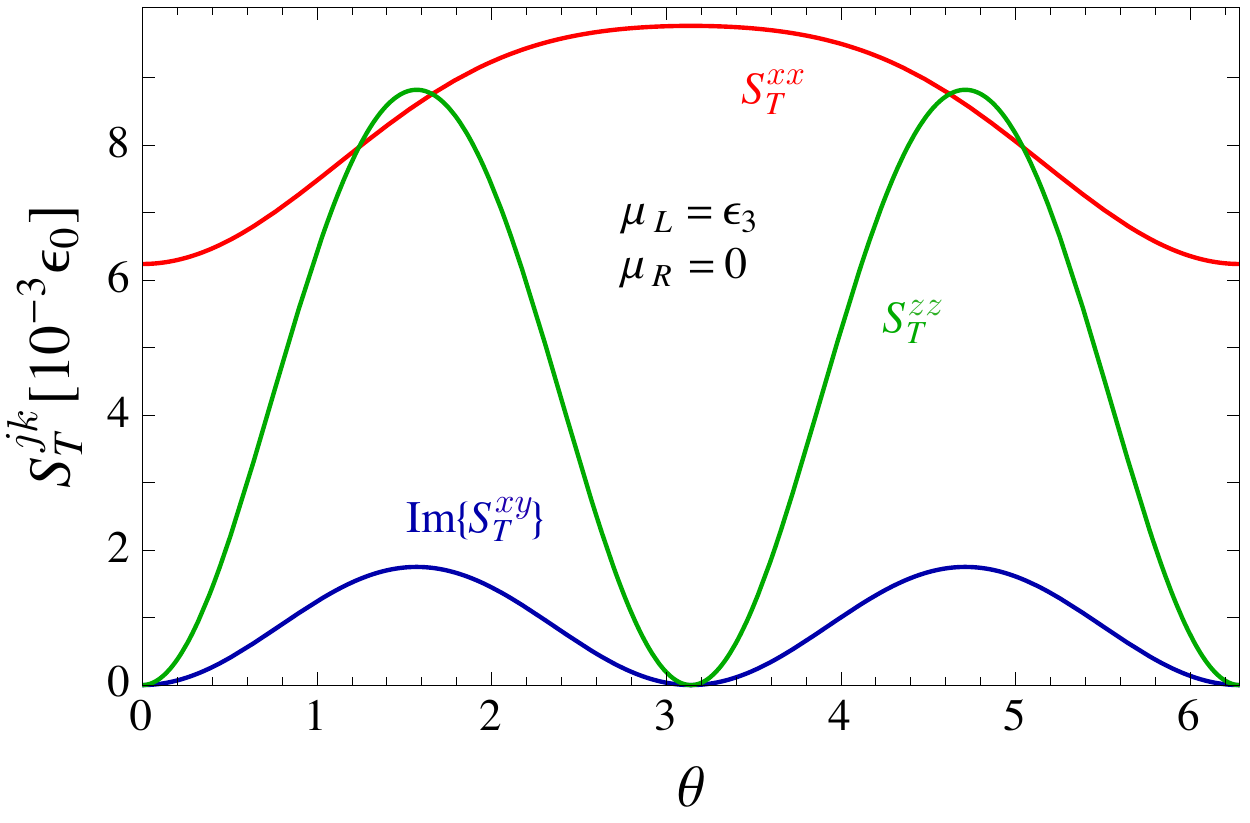}
	\caption{(Color online) Spin-torque shot-noise components as functions of the tilt angle $\theta$ for $\mu_{L}=\epsilon_{3}$, $\mu_{R}=0$. All plots are obtained at zero temperature, with $\vec B=B\vec e_{z}$, $\Gamma=0.05\,\epsilon_{0}$, and $\Gamma_{L}=\Gamma_{R}=\Gamma/2$. The other parameters are $\omega_{L}=0.5\,\epsilon_{0}$, $J=0.01\,\epsilon_{0}$, and $S=100$. }\label{fig: Noisetheta}
\end{figure}
For $\omega=\epsilon_{0}$ corresponding to $\mu_{R}=\epsilon_{1}$ and $\mu_{R}=\epsilon_{4}-2\omega_{L}$, both the elastic and spin-flip tunneling events involving the absorption of energy of one quantum $\omega_L$ contribute with a dip, while the spin-flip processes involving the absorption of an energy equal to $2\omega_L$ contribute with a peak to the component $S^{xx}_{T}$. For $\omega_{L}=2\,\epsilon_{0}$ and $\omega_{L}=3\,\epsilon_{0}$ corresponding to $\mu_{L}=\epsilon_{2}$ and $\mu_{R}=\epsilon_{3}$, both elastic and spin-flip processes with the absorption of an energy equal to $\omega_L$ contribute with a step, while the inelastic processes involving the absorption of an energy $2\omega_{L}$ give negligible contribution to $S^{xx}_{T}$. The component $S^{zz}_{T}$ shows dips at these two points, since here the dominant contribution comes from inelastic tunneling spin-flip events. The component $S^{zz}_{T}$ is an even function of $\omega_L$, while ${\rm Im}\{S^{xy}_{T}\}$ is an odd function of $\omega_{L}$. The spin-torque noise $S^{xx}_{T}$ is an even function of $\omega_L$ for $\theta=\pi/2$.

The spin-torque noise components as functions of $\theta$ for
$\mu_{L}=\epsilon_{3}$ and $\mu_{R}=0$ at zero temperature are shown
in Fig.~\ref{fig: Noisetheta}. The magnitudes and the appearance of
the spin-torque noise components at resonance  
energies $\epsilon_i$ can be controlled by $\theta$, since it
influences the polarization of the spin current. Here we see that both
$S^{zz}_{T}$ and ${\rm Im}\{S^{xy}_{T}\}$ are zero for $\theta=0$ and
$\theta=\pi$, as the molecular spin is static and its magnitude is
constant along $z$ direction in both cases. These torque-noise
components take their maximum values for $\theta=\pi/2$, where both
elastic and inelastic tunneling contributions are maximal. The
component $S^{xx}_{T}$ takes its minimum value for $\theta=0$ and its
maximum value for $\theta=\pi$, with only elastic tunneling
contributions in both cases. For $\theta=\pi/2$, the inelastic
tunneling events make a maximal contribution while energy conserving
processes give minimal contribution to $S^{xx}_{T}$.

\section{Conclusions}\label{sec: conclusions3ndpaper}

In this article, we studied theoretically the noise of charge and spin transport through a small junction, consisting of a single molecular orbital in the presence of a molecular spin precessing with Larmor frequency $\omega_L$ in a constant magnetic field. The orbital is connected to two Fermi leads.
We used the Keldysh nonequilibrium Green's functions method to derive the noise components of charge and spin currents and spin-transfer torque.

Then, we analyzed the shot noise of charge current and observed characteristics that differ from the ones in the current. In the noise power, we observed diplike features which we attribute to inelastic processes, due to the molecular spin precession, leading to the quantum-interference effect between correlated transport channels.

Since the inelastic tunneling processes lead to a spin-transfer torque
acting on the molecular spin, we have also investigated the spin-torque noise components contributed by these processes, involving the change of energy by an energy quantum $\omega_{L}$.  The spin-torque noise components are driven by both the bias voltage and the molecular spin precession. The in-plane noise components $S^{xx}_{T}$ and $S^{yy}_{T}$ are also contributed by the processes involving the absorption of an energy equal to $2\omega_{L}$. We obtained the relation between ${\rm Im}\{S^{xy}_{T}\}$ and the Gilbert damping coefficient $\alpha$ at arbitrary temperature.

Taking into account that the noise of charge and spin transport can be controlled by the parameters such as bias voltage and external magnetic field, 
our results might be useful in molecular electronics and spintronics.
The experimental observation of the predicted noise properties might be a challenging task due to complicated tunnelling processes through molecular magnets.
Finding a way to control the spin states of single-molecule magnets in tunnel junctions could be one of the future tasks.

\begin{acknowledgments}
	We would like to thank Fei Xu for useful discussions. We
        gratefully acknowledge the financial support from the Deutsche
        Forschungsgemeinschaft through the SFB 767 \textit{Controlled
          Nanosystems}, the Center of Applied Photonics, the DAAD
        through a STIBET scholarship, and an ERC
        Advanced Grant \textit{UltraPhase} of Alfred Leitenstorfer. 
\end{acknowledgments}

\begin{widetext}

	\appendix*
		
    \section*{Appendix: Formal expression for the nonsymmetrized noise}
     \renewcommand{\theequation}{A\arabic{equation}}
    % redefine the command that creates the equation no.
    \setcounter{equation}{0}  % reset counter 
   % \section*{Appendix: Expressions for spin-current components}  % use *-form to suppress numbering
   
   Here, we present the derivation of the formal expression for the nonsymmetrized noise $S^{\nu\mu}_{\xi\zeta}(t,t')$. The correlation functions $S^{\sigma\sigma',\lambda\eta}_{\xi\zeta}(t,t')$, introduced in Eq.~(\ref{eq: 5.6}), can be expressed by means of the Wick's theorem \cite{Walecka} as
   \begin{align}
   	S^{\sigma\sigma',\lambda\eta}_{\xi\zeta}(t,t')&=
   	\sum_{kk'}[V_{k\xi}V_{k'\zeta}G^{>}_{\sigma',k'\lambda\zeta}(t,t')G^{<}_{\eta,k\sigma\xi}(t',t)\nonumber\\
   	&-V_{k\xi}V^{*}_{k'\zeta}G^{>}_{\sigma'\lambda}(t,t')G^{<}_{k'\eta\zeta,k\sigma\xi}(t',t)\nonumber\\
   	&-V^{*}_{k\xi}V_{k'\zeta}G^{>}_{k\sigma'\xi,k'\lambda\zeta}(t,t')G^{<}_{\eta\sigma}(t',t)\nonumber\\
   	&+V^{*}_{k\xi}V^{*}_{k'\zeta}G^{>}_{k\sigma'\xi,\lambda}(t,t')G^{<}_{k'\eta\zeta,\sigma}(t',t)],\label{eq: 5.7}
   \end{align}
   with the mixed Green's functions defined, using units in which $\hbar=e=1$, as
   \begin{align}
   	G^{<}_{\eta,k\sigma\xi}(t,t')&=i\langle\hat{c}^{\dag}_{k\sigma\xi}(t')\hat{d}_{\eta}(t)\rangle,\\
   	G^{>}_{\sigma',k'\lambda\zeta}(t,t')&=-i\langle\hat{d}_{\sigma'}(t)\hat{c}^{\dag}_{k'\lambda\zeta}(t')\rangle,
   \end{align}
   while Green's functions $G^{<}_{k\sigma\xi,\eta}(t,t')=-[G^{<}_{\eta,k\sigma\xi}(t',t)]^{*}$ and $G^{>}_{k'\lambda\zeta,\sigma'}(t,t')=-[G^{>}_{\sigma',k'\lambda\zeta}(t',t)]^{*}$.
   The Green's functions of the leads and the central region are defined as
   \begin{align}
   	G^{<}_{k\sigma\xi,k'\sigma'\zeta}(t,t')&=i\langle\hat{c}^{\dag}_{k'\sigma'\zeta}(t')\hat{c}_{k\sigma\xi}(t)\rangle,\\
   	G^{>}_{k\sigma\xi,k'\sigma'\zeta}(t,t')&=-i\langle\hat{c}_{k\sigma\xi}(t)\hat{c}^{\dag}_{k'\sigma'\zeta}(t')\rangle,\\
   	G^{<}_{\sigma\sigma'}(t,t')&=i\langle\hat{d}^{\dag}_{\sigma'}(t')\hat{d}_{\sigma}(t)\rangle,\label{eq: 5.8m}\\
   	G^{>}_{\sigma\sigma'}(t,t')&=-i\langle\hat{d}_{\sigma}(t)\hat{d}^{\dag}_{\sigma'}(t')\rangle,\label{eq: 5.9m}\\
   	G^{r,a}_{\sigma\sigma'}(t,t')&=\mp i\theta(\pm t\mp t')\langle\{\hat{d}_{\sigma}(t),\hat{d}^{\dag}_{\sigma'}(t')\}
   	\rangle.\label{eq: 5.10m}	
   \end{align}	
   Since the self-energies originating from the coupling between the electronic level and the lead $\xi$ are diagonal in the electron spin space, their entries can be written as $\Sigma^{<,>,r,a}_{\xi}(t,t')=\sum_{k}V_{k\xi}g_{k\xi}^{<,>,r,a}(t,t')V^{*}_{k\xi}$, where $g^{<,>,r,a}(t,t')$ are the Green's functions of the free electrons in lead $\xi$.  
   Applying Langreth analytical continuation rules,\cite{Langreth1976} Eq.~(\ref{eq: 5.7}) transforms into
   \begin{align}
   	S^{\sigma\sigma',\lambda\eta}_{\xi\zeta}(t,t')=&\int dt_{1}\int dt_{2}\nonumber\\
   	\times&\big\{[G^{r}_{\sigma'\lambda}(t,t_{1})\Sigma^{>}_{\zeta}(t_{1},t')+G^{>}_{\sigma'\lambda}(t,t_{1})\Sigma^{a}_{\zeta}(t_{1},t')]
   	[G^{r}_{\eta\sigma}(t',t_{2})\Sigma^{<}_{\xi}(t_{2},t)+G^{<}_{\eta\sigma}(t',t_{2})\Sigma^{a}_{\xi}(t_{2},t)]\nonumber\\
   	+&[\Sigma^{>}_{\xi}(t,t_{1})G^{a}_{\sigma'\lambda}(t_{1},t')+\Sigma^{r}_{\xi}(t,t_{1})G^{>}_{\sigma'\lambda}(t_{1},t')]
   	[\Sigma^{<}_{\zeta}(t',t_{2})G^{a}_{\eta\sigma}(t_{2},t)+\Sigma^{r}_{\zeta}(t',t_{2})G^{<}_{\eta\sigma}(t_{2},t)]\nonumber\\
   	-&G^{>}_{\sigma'\lambda}(t,t')[\Sigma^{r}_{\zeta}(t',t_{1})G^{r}_{\eta\sigma}(t_{1},t_{2})\Sigma^{<}_{\xi}(t_{2},t)
   	+\Sigma^{<}_{\zeta}(t',t_{1})G^{a}_{\eta\sigma}(t_{1},t_{2})\Sigma^{a}_{\xi}(t_{2},t)\nonumber\\
   	+&\Sigma^{r}_{\zeta}(t',t_{1})G^{<}_{\eta\sigma}(t_{1},t_{2})\Sigma^{a}_{\xi}(t_{2},t)]
   	-[\Sigma^{r}_{\xi}(t,t_{1})G^{r}_{\sigma'\lambda}(t_{1},t_{2})\Sigma^{>}_{\zeta}(t_{2},t')\nonumber\\
   	+&\Sigma^{>}_{\xi}(t,t_{1})G^{a}_{\sigma'\lambda}(t_{1},t_{2})\Sigma^{a}_{\zeta}(t_{2},t')
   	+\Sigma^{r}_{\xi}(t,t_{1})G^{>}_{\sigma'\lambda}(t_{1},t_{2})\Sigma^{a}_{\zeta}(t_{2},t')]G^{<}_{\eta\sigma}(t',t)\big\}\nonumber\\
   	\label{eq: 5.14}-&\delta_{\xi\zeta}[\delta_{\eta\sigma}G^{>}_{\sigma'\lambda}(t,t')\Sigma^{<}_{\xi}(t',t)
   	+\delta_{\sigma'\lambda}\Sigma^{>}_{\xi}(t,t')G^{<}_{\eta\sigma}(t',t)].
   \end{align} 
Finally, using Eqs.~(\ref{eq: 5.6}) and (\ref{eq: 5.14}), the obtained formal expression for the nonsymmetrized noise of charge current\cite{JauhoBook,FengMaciejko075302} and spin currents in standard coordinates $t$ and $t'$ can be written as 
\begin{align}
	S^{\nu\mu}_{\xi\zeta}(t,t')&=-\frac{q_{\nu}q_{\mu}}{\hbar^2}{\rm Tr}\Big\{\int dt_{1}\int dt_{2}\nonumber\\
	&\times\big\{\hat{\sigma}_{\nu}[\hat{G}^{r}(t,t_{1})\hat{\Sigma}^{>}_{\zeta}(t_{1},t')+\hat{G}^{>}(t,t_{1})\hat{\Sigma}^{a}_{\zeta}(t_{1},t')]\hat{\sigma}_{\mu}[\hat{G}^{r}(t',t_{2})\hat{\Sigma}^{<}_{\xi}(t_{2},t)+\hat{G}^{<}(t',t_{2})\hat{\Sigma}^{a}_{\xi}(t_{2},t)]\nonumber\\
	&+\hat{\sigma}_{\nu}[\hat{\Sigma}^{>}_{\xi}(t,t_{1})\hat{G}^{a}(t_{1},t')+\hat{\Sigma}^{r}_{\xi}(t,t_{1})\hat{G}^{>}(t_{1},t')]\nonumber\hat{\sigma}_{\mu}[\hat{\Sigma}^{<}_{\zeta}(t',t_{2})\hat{G}^{a}(t_{2},t)+\hat{\Sigma}^{r}_{\zeta}(t',t_{2})\hat{G}^{<}(t_{2},t)]\nonumber\\
	&-\hat{\sigma}_{\nu}\hat{G}^{>}(t,t')\hat{\sigma}_{\mu}[\hat{\Sigma}^{r}_{\zeta}(t',t_{1})\hat{G}^{r}(t_{1},t_{2})\hat{\Sigma}^{<}_{\xi}(t_{2},t)+\hat{\Sigma}^{<}_{\zeta}(t',t_{1})\hat{G}^{a}(t_{1},t_{2})\hat{\Sigma}^{a}_{\xi}(t_{2},t)+\hat{\Sigma}^{r}_{\zeta}(t',t_{1})\hat{G}^{<}(t_{1},t_{2})\hat{\Sigma}^{a}_{\xi}(t_{2},t)]\nonumber\\
	&-\hat{\sigma}_{\nu}[\hat{\Sigma}^{r}_{\xi}(t,t_{1})\hat{G}^{r}(t_{1},t_{2})\hat{\Sigma}^{>}_{\zeta}(t_{2},t')+\hat{\Sigma}^{>}_{\xi}(t,t_{1})\hat{G}^{a}(t_{1},t_{2})\hat{\Sigma}^{a}_{\zeta}(t_{2},t')+\hat{\Sigma}^{r}_{\xi}(t,t_{1})\hat{G}^{>}(t_{1},t_{2})\hat{\Sigma}^{a}_{\zeta}(t_{2},t')]\hat{\sigma}_{\mu}\hat{G}^{<}(t',t)\big\}\nonumber\\
	&-\delta_{\xi\zeta}\hat{\sigma}_{\nu}[\hat{G}^{>}(t,t')\hat{\sigma}_{\mu}\hat{\Sigma}^{<}_{\xi}(t',t)+\hat{\Sigma}^{>}_{\xi}(t,t')\hat{\sigma}_{\mu}\hat{G}^{<}(t',t)]\Big\},\label{eq: 5.7k}
\end{align}
where $\rm{Tr}$ denotes the trace in the electronic spin space.	
\end{widetext}

\end{document}